\documentclass[preprint]{aastex}
\usepackage[]{graphicx}

\begin{document}

\title{The Spectral Evolution of Convective Mixing White Dwarfs, the non-DA Gap, and White Dwarf Cosmochronology}
\author{Eugene Y. Chen\altaffilmark{1} \& Brad M. S. Hansen\altaffilmark{2}
}
\altaffiltext{1}{Department of Astronomy, The University of Texas at Austin, Austin, TX 78712, USA; eyc@mail.utexas.edu}
\altaffiltext{2}{Department of Physics and Astronomy, University of California Los Angeles, Los Angeles, CA 90024, USA; hansen@astro.ucla.edu}
\shortauthors{Chen \& Hansen }
\shorttitle{Convective mixing and non-DA gap}

\begin{abstract}
The spectral distribution of field white dwarfs shows a feature called the ``non-DA gap''. As defined by Bergeron et al., this is a temperature range ($5100K$--$6100K$) where relatively few non-DA stars are found, even though
such stars are abundant on either side of the gap.  It is usually viewed as an indication that a significant fraction of 
white dwarfs switch their atmospheric compositions back and forth between hydrogen-rich and helium-rich as they cool.  In this paper, we present a Monte Carlo model of the Galactic disk white dwarf population, based on the spectral evolution model of Chen and Hansen.  
We find that the non-DA gap emerges naturally, even though our model only allows white dwarf atmospheres to evolve monotonically from hydrogen-rich to helium-rich through convective mixing.  We conclude by discussing the effects of convective mixing on the white dwarf luminosity function and the use thereof for Cosmochronology.
\end{abstract}


\section{The non-DA gap as a complication of WD Cosmochronology}
\label{sec:nonDAgapIntro}
White dwarfs (WDs) are the endpoint of stellar evolution for low mass stars. 
Their spectra can be divided into two broad categories:  the DA stars ($\sim75\%$), whose atmospheric composition is observed to be hydrogen-dominated, and the non-DA stars ($\sim25\%$), whose atmospheric composition is helium-dominated.  The evolution of WDs is mainly driven by the cooling of the degenerate core.  As a result, the effective temperature ($T_{eff}$), or equivalently, the luminosity of a WD can serve as a cosmic clock, indicating the time passed since its formation \citep{2001PASP..113..409F}.  Analytical and numerical models of WD cooling have been made (e.g., \citet{1952MNRAS.112..583M,1975ApJ...200..306L,1992ApJ...386..539W,1997ApJ...477..313A,1999ApJ...520..680H,2000ApJ...543..216C,2000ApJ...544.1036S,
2008ApJ...677..473G,2010ApJ...717..183R}), allowing the use of white dwarf luminosity function (WDLF) to derive the age of Galactic disk \citep{1988ApJ...332..891L,1987ApJ...315L..77W,1992ApJ...386..539W} and other stellar populations
 \citep{2002ApJ...574L.155H,2005ApJ...624L..45B,2007ApJ...671..380H,2009ApJ...697..965B}.  The practice is known as WD Cosmochronology \citep{1987ApJ...315L..77W}.

However, there are still many complications that impede WD Cosmochronology.
Among them are the observations of \citet{1997ApJS..108..339B,2001ApJS..133..413B}.  In their work, photometric and spectroscopic observations of 152 \emph{cool} ($4000K<T_{eff}<12000K$) WDs that belong to the local Galactic disk are analyzed with state-of-the-art model atmospheres of pure hydrogen, pure helium and H/He mixtures.  It was shown that the ratio of observed helium-rich stars to hydrogen-rich stars is greatly reduced in the temperature range $5100K$--$6100K$ and enhanced both above and below this temperature range (see Figure~\ref{f1}).  This temperature range was consequently termed the ``non-DA gap'' to indicate that few non-DA stars are found within.  Furthermore, some so-called ``peculiar stars'', whose spectral distributions are better reproduced with hydrogen model atmospheres but which show no $H_\alpha$ lines, are found in the temperature range $\sim 6000K$ to $\sim 8500K$.  The results seemed to portray the following picture of WD spectral evolution as it cools from $T_{eff}=12000K$ to $T_{eff}=4000K$:  First, a significant fraction of the DA stars (which are the majority of all WDs) switch their atmospheric composition from hydrogen-rich to helium-rich in the temperature range $12000K\sim 8500K$, causing a steady increase of non-DA ratio (with respect to the decrease of $T_{eff}$).  Later, some helium stars start to evolve into peculiar stars (at $\sim8500K$), and then into hydrogen-rich stars (at $\sim6100K$).  The stars remain hydrogen-rich in the ``non-DA gap'' ($6100K\sim 5100K$) and transform into helium-rich WDs again at $\sim5100K$.  The observations introduced an uncertainty in WD Cosmochronology because atmospheric composition significantly affects cooling rate at lower temperatures (e.g., \citet{1999ApJ...520..680H}), yet, all of the previous works on theoretical WDLF were done with only models of fixed atmospheric composition.

\citet{1997ApJS..108..339B} proposed an accretion--convective mixing\footnote{We refer the reader to section~\ref{sec:cMixIntro} for a review of convective mixing.}
model to account for the observations.  In their model, hydrogen-rich inter-stellar medium is first accreted onto the surface of a helium WD at temperatures higher than $6100K$ without being spectroscopically visible, due to the quenching effects
of the high photospheric pressure that characterizes these atmospheres.  Such a WD will appear as a peculiar star, until the
hydrogen concentration reaches the level at which $H^{-}$ starts to recombine.  At this point, a runaway process will occur and turn the star into a DA star at $\sim 6100K$.  The spectral type of the star remains DA as it cools from $6100K$ to $5100K$.  Finally, at $5100K$, it is transformed into a helium-rich star again through convective mixing.  However, this proposal was later rejected by \citet{1999ApJ...517..901M} as they claim no models
in their grids were found to have $H_\alpha$ lines quenched while retaining the spectral energy distribution of a pure hydrogen atmosphere.

An alternative explanation was given in \citet{1999ApJ...520..680H}, where the non-DA gap was suggested to be the consequence of the different cooling rates of hydrogen-rich and helium-rich WDs.  However, it was later pointed out in \citet{2001ApJS..133..413B} that such a proposal cannot explain the enhancement of the non-DA ratio at temperatures above the blue edge of the gap.  As a consequence, there is still no convincing explanation for the non-DA gap and its properties.

\section{Convective Mixing as a Mechanism of Spectral Evolution}
\label{sec:cMixIntro}
It has been established that elemental diffusion is efficient in the WD atmosphere \citep{1984ApJ...278..769M}.  In the absence of other mechanisms, a WD with only small amounts of hydrogen therefore possesses a photosphere close to pure hydrogen.  The leading mechanism to transform a hydrogen-rich (photosphere) WD to a helium-rich WD is through a process called convective mixing, first proposed by \citet{1972ApJ...177..723S}.  The idea is as follows:  The surface convection zone of a hydrogen WD grows as $T_{eff}$ decreases.  For WDs with intermediate amount of hydrogen (i.e., the hydrogen layer is optically thick but less massive than $\sim10^{-6}M_\sun$), the convection zone will eventually reach the underlying helium layer.  The helium will thus be brought to the surface through convective motion, changing the photospheric composition.

This mechanism has been investigated extensively, and a number of outcomes have been proposed.
\citet{1972ApJ...177..723S} and \citet{1973A&A....27..307B} suggested that a temperature increase accompanies convective mixing. However, the evolution model of \citet{1976A&A....52..415K} appears to show no such behavior.  Besides, while most groups conjectured (based on the size of convection zones in pure hydrogen and helium WDs)
that the composition of the resultant photosphere would be close to pure helium, \citet{1990ApJ...351L..21B}, based on observational results, suggested that the convective mixing might result in an H/He mixture whose opacity is still dominated by hydrogen.  The work of \citet{2011MNRAS.413.2827C} succeeded in unifying all of the suggested outcome into one framework.  Using WD mass ($M_{WD}$) and total hydrogen mass ($m_H$) as the input parameters, a set of self-consistent, quantitative cooling curves and chemical evolution curves were derived (see, e.g., Figure~\ref{f2}).  It was concluded that the temperature increase which accompanies convective mixing only occurs when the post-mixing envelope is convectively coupled to the degenerate core, and that the post-mixing photosphere hydrogen fraction ($X_{surf}$) is a function of both $M_{WD}$ and $m_H$.  It has been shown that massive WDs with a higher amount of hydrogen are more likely to stay hydrogen-rich after convective mixing.

The convective mixing mechanism has also been deployed in the aforementioned model of \citet{1997ApJS..108..339B}, in which the authors seemed to implicitly assume no temperature increase and a post-mixing photosphere which can be treated as pure helium.  Based on the results of \citet{2011MNRAS.413.2827C}, these assumptions require some modifications.  The $T_{eff}$ should increase upon mixing because the WD envelopes are already convectively coupled to the degenerate core.  Another feature of the convective mixing WDs that Bergeron et al. did not fully exploit is their faster, post-mixing cooling rate (cf. Figure~\ref{f2}).

In the rest of this paper, we will demonstrate that the non-DA gap naturally emerges from
the improved treatment of convective mixing, without additional mechanisms. In order to do
this, we will construct a model realization of the Galactic disk WD population based
on the recent determinations of Galactic parameters.

\section{A Monte Carlo realization of the Galactic white dwarf population}
\label{sec:mcSample}
In this section, we will describe the construction of a Monte Carlo sample of Galactic WDs which includes the treatment of convective mixing of
\citet{2011MNRAS.413.2827C}.  The purpose is to see if such inclusion could account for the non-DA gap of
\citet{1997ApJS..108..339B,2001ApJS..133..413B}.  Specifically, the non-DA to total WD
ratio as a function of $T_{eff}$ will be calculated at
a specific Galactic disk age $t_d$.  We will also compare the WDLF synthesized from this sample to some other theoretical and observational WDLF and discuss the cosmological implications.

The Monte Carlo sample is defined by an array ($\sim 10^6$ entries, one entry per star) that specifies the physical properties of each WD (or its progenitor, if the star is too young to have evolved).  Two random numbers, stellar formation time ($t_b$) and main sequence stellar mass ($M_{MS}$), are drawn for each star.  As an approximation, we take the star formation rate of our Galaxy to be constant and generate $t_b$ to be uniformly distributed between $0$ and $t_d$.  $M_{MS}$ is generated to conform with the initial mass function
of \citet{2001MNRAS.322..231K} with an upper limit of $8M_\sun$ (stars with masses $>8M_\sun$ are not statistically important).  A pre-WD lifespan $t_{MS}$ is assigned to each star based on its $M_{MS}$,
following the formula described in \citet{2000MNRAS.315..543H}.  A comparison between $t_d$ and $(t_b+t_{MS})$ is then made:  If $t_d<(t_b+t_{MS})$, the star is not yet a WD and is
discarded;  If $t_d>(t_b+t_{MS})$, we further calculate $M_{WD}$ using the relation established by \citet{2012ApJ...746..144Z}.  We define the difference between the two quantities as the cooling age $t_c$, i.e., $t_c\equiv t_d-t_b-t_{MS}$.

All physical quantities of a WD can be derived once $M_{WD}$, $t_c$, and a WD evolution model grid are specified.  Three WD evolution grids are available in this work:  the pure helium WD models with $m_H$ being equal to zero, the thin hydrogen layer WD models with $m_H=10^{-9}M_\sun$ and the thick hydrogen layer WD models with a hydrogen fractional mass $q(H)\equiv m_H/M_{WD}=10^{-4}$.  All WDs in our sample have helium fractional mass $q(He)=10^{-2}$.  The pure helium WD models and thick hydrogen layer
WD models are described in \citet{1999ApJ...520..680H} and the thin hydrogen layer WD models are described in \citet{2011MNRAS.413.2827C}.
The $m_H$ value we chose for the thin hydrogen layer stars, $10^{-9}M_\sun$, is a value consistent with asteroseismology studies.  The hydrogen fractional mass we chose for the thick hydrogen layer stars and the helium fractional mass we chose for all WDs are the maximum amount that can survive the pre-WD evolutionary phases and is considered as the ``standard thick layer model'' \citep{1984ApJ...283..787M,1990ARA&A..28..139D,2001PASP..113..409F}.  It is a simplification to account for all WDs with only the mentioned three model grids.  In any real WD samples, either $m_H$ or $q(H)$ should more or less follow a continuous distribution.  The $m_H$ or $q_H$ values we adopt in our work can be viewed as a set of characteristic values, i.e., where the distribution is peaked.

In the following paragraphs, we show the results of one of our simulations with $t_d=9.5\, Gyr$.
We assume 15\% of the WDs are pure helium WDs, 32\% of the WDs are thin hydrogen layer WDs and 53\% of the WDs are thick hydrogen layer
WDs.  The choice of disk age, $t_d=9.5\, Gyr$, is consistent with \citet{1998ApJ...497..294L}.  The WD type fractions are chosen to
match the spectral type distribution from observations (cf. Figure~\ref{f4}).  However, it should be kept in mind that our main purpose is to show the qualitative consistency between our model and observations.  No attempts were made to accurately fit for either the Galactic age or the percentage of convective mixing WDs.

\begin{figure}
\includegraphics[width=0.8\textwidth]{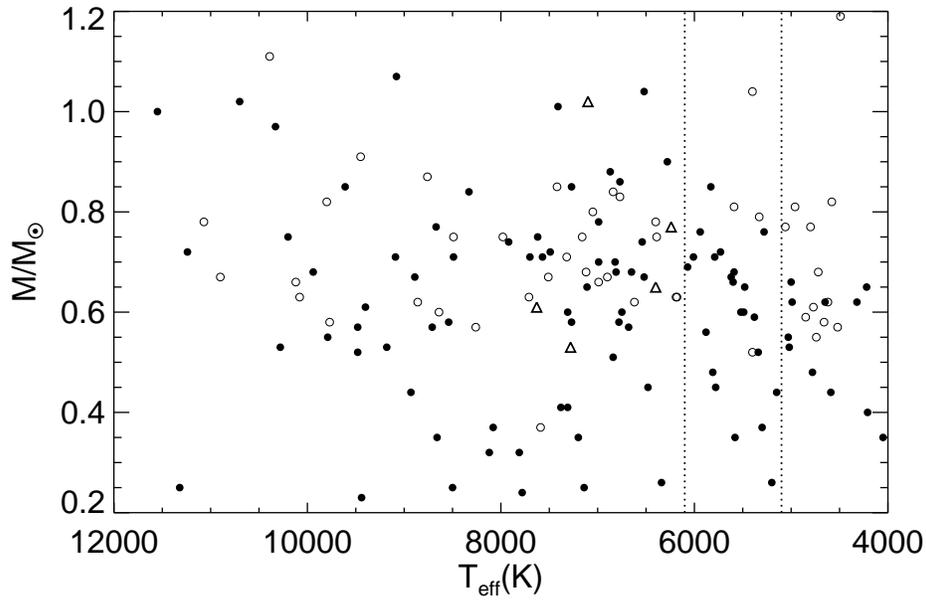}
\caption{A simplified version of Figure~21 of \citet{2001ApJS..133..413B}.  Filled Circle: H-rich star, Hollow Circle: He-rich star,
Triangle: peculiar star.  The main atmospheric composition is taken from their Table~2.  Complications such as $DQ$ stars, $C_2H$ stars\dots, etc. are not presented.}
\label{f1}
\end{figure}

Figure~\ref{f3} shows the distribution of a representative Monte Carlo WD population in the $T_{eff}$--$M_{WD}$ space and $T_{eff}$--$X_{surf}$ space.  We have reduced the number of stars in the sample to avoid the plot being over-congested (we used the full amount of numerical stars in making Figure~\ref{f4} and Figure~\ref{f5} and confirmed that the results have already converged with respect to sample size).  Following \citet{1999ApJ...517..901M}, WDs with $X_{surf}<10^{-3}$ are classified as helium-rich stars and those with $X_{surf}>10^{-2}M_\sun$ are classified as hydrogen-rich stars.  Interestingly, a small population of intermediate $X_{surf}$ stars ($10^{-3}<X_{surf}<10^{-2}$) are found to cluster around $\sim8500K$ to $\sim6000K$, which is the location of the peculiar stars of \citet{1997ApJS..108..339B,2001ApJS..133..413B}.  Therefore, they seem to be natural \emph{candidates} for the observed population of peculiar stars.  These stars are evolved from intermediate massive ($0.69M_\sun\lesssim M_{WD}\lesssim 0.87M_\sun$) thin hydrogen layer WDs.  The higher stellar masses of these WDs make their convection zones shallower (compare to the majority of WDs which are less massive) and consequently leads to the intermediate hydrogen fraction that results upon convective mixing (those with even higher masses would remain hydrogen-rich stars).  At later times, most of them evolve into true helium-rich stars due to the deepening of their convection zones, hence they are \emph{less likely} to be found below $6000K$.  We would like to emphasize that our plot (upper panel, the $T_{eff}$--$M_{WD}$ plot) is not a simulation of Figure~21 of \citet{2001ApJS..133..413B} (i.e., our Figure~\ref{f1}) and is intrinsically different from it in the following two respects.  First, the ``stellar density distribution'' is different (the stars in Figure~\ref{f3} are aggregated in the lower-right region whereas the stars in Figure~\ref{f1} are more uniformly distributed), because our plot is made with the whole sample instead of the \emph{detected} sample.  In order to properly compare observational results with our numerical work, information on detectability needs to be incorporated (cf. the next two paragraphs).
Secondly, our numerical sample does not have any WDs less massive than $\sim0.53M_\sun$, because the progenitor of these low mass WDs would still be on the main sequence track, according to our $M_{MS}$--$M_{WD}$ relation.  The low mass WDs in Figure~\ref{f1} could be evolved from, e.g., binary systems, and are not accounted for in our model.

While the non-DA gap is \emph{qualitatively} apparent in \citet{2001ApJS..133..413B}, the most up-to-date and \emph{quantitative} observational data on the non-DA ratio of cold WDs is in the works of \citet{2008ApJ...672.1144T} and \citet{1998ApJ...497..294L}.  In Figure~\ref{f4}, we will directly compare our numerical results with these observations.  The data in \citet{2001ApJS..133..413B} have a larger sample size, however, detectability information, e.g., the $V_{max}$ correction \citep{1975ApJ...202...22S}, is absent.  Therefore, a quantitative comparison with our work cannot be made at this point.

Two sets of observational data are plotted in Figure~\ref{f4} in comparison with our numerical results (solid line).  The two dotted lines
are the upper and lower limits of the non-DA to total WD ratio derived from Figure~11 of \citet{2008ApJ...672.1144T}.  The filled circles with error bars are compiled by us from the results of \citet{1998ApJ...497..294L}: instead of binning their WD sample
(which is an updated determination of the 43 faintest WDs presented in \citet{1988ApJ...332..891L}) in luminosity, we binned the sample (with $V_{max}$ correction) in $T_{eff}$.  The binning was done for the DA (hydrogen-rich) stars and the non-DA (helium-rich) stars respectively.  In each bin, we calculate the upper limit of the non-DA to total WD ratio by dividing the upper limit of the non-DA density to the lower limit of total WD density (while fixing the non-DA density at its upper limit), and the lower limit of the ratio by dividing the lower limit of the non-DA density to the upper limit of the total WD density (while fixing the non-DA density at its lower limit).  All of the results are plotted in Figure~\ref{f4} except that corresponds to the bin $8000K$--$9000K$, where the result and statistical error are of the same size (because there is only one star in the sample).
Our numerical result (solid line) is generated by directly dividing the number of non-DAs to the number of all WDs in each $T_{eff}$ bin (1000K in size) of our Monte Carlo sample.  We can see that a non-DA gap naturally emerges in the range $5000K$--$6000K$ as a consequence of convective mixing and the resulting faster cooling rates of post-mixing WDs.

In Figure~\ref{f5}, we examine how this qualitative change in the evolution of individual stars is reflected in the luminosity function of the population as a whole.  We compile the WDLF
from our Monte Carlo sample which includes convective mixing WDs (CM-WDLF hereafter, solid line), together with that from a sample composed solely of thick hydrogen layer WDs (TH-WDLF hereafter, dash-dotted line) and compare to three different sources of observational WDLF.  The three different sources are \citet{1988ApJ...332..891L} (LDM hereafter, for $L>10^{-3.2}L_\sun$), \citet{1998ApJ...497..294L} (LRB hereafter, for $L<10^{-3.2}L_\sun$), and the results from Sloan Digital Sky Survey \citep{2006AJ....131..571H} (SDSS hereafter, for the entire range of luminosity).

The CM-WDLF starts to deviate from the TH-WDLF below $L=10^{-3}L_\sun$.  The value of the CM-WDLF is larger than that of the TH-WDLF between $L=10^{-3}L_\sun$ and $L=10^{-3.5}L_\sun$, due to the temperature increase that accompanies convective mixing.  Below $L=10^{-3.5}L_\sun$, however, the situation inverts, because the convective mixing WDs (and the pure helium WDs) have faster cooling rates, resulting in a flatter,
less peaked WDLF.  The faster cooling rates also enable the WDLF to extend to a fainter luminosity.
The observations from LDM and LRB are not sufficiently densely sampled to distinguish between the two different theoretical shapes, but the SDSS sample may have that capability once its properties are more accurately determined (e.g., when the parallax of each star are obtained to determine the distances; currently, the SDSS WDLF are derived with photometric distance and an idealized assumption of $\log g=8.0$ for all WDs is made).

\section{Conclusions and discussions}
\label{sec:conclusions}

We have demonstrated that the non-DA gap can be reproduced in a population model by exploiting the spectral evolution model investigated in \citet{2011MNRAS.413.2827C}.  The non-DA gap can be formed by the combined effects of convective mixing and the higher cooling rates of the post-mixing WDs.  An evolution model that transforms DC stars into DA stars, e.g., the one outlined in \citet{1997ApJS..108..339B}, is not needed to produce the non-DA gap.

\begin{figure}
\includegraphics[width=0.8\textwidth]{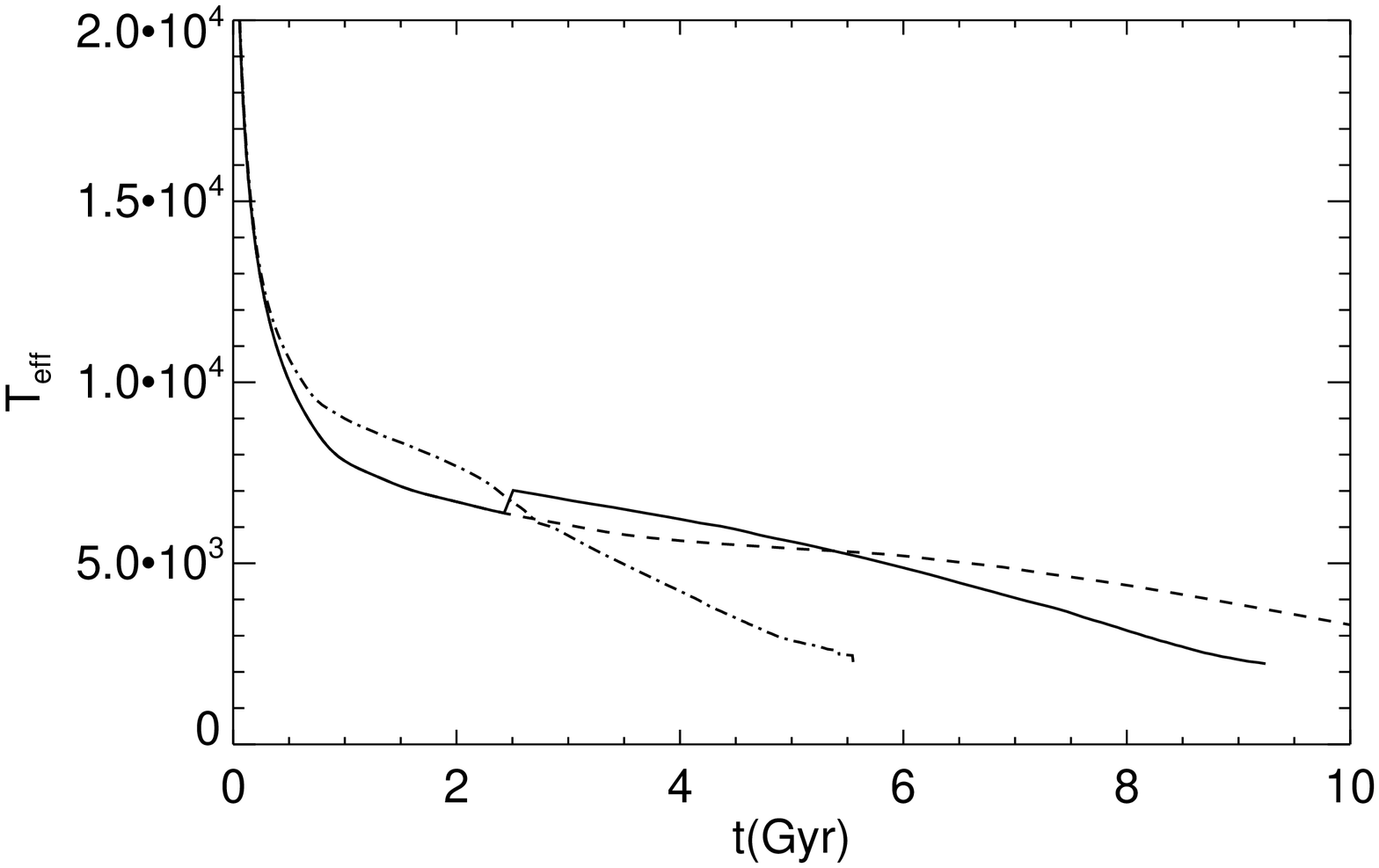}
\includegraphics[width=0.8\textwidth]{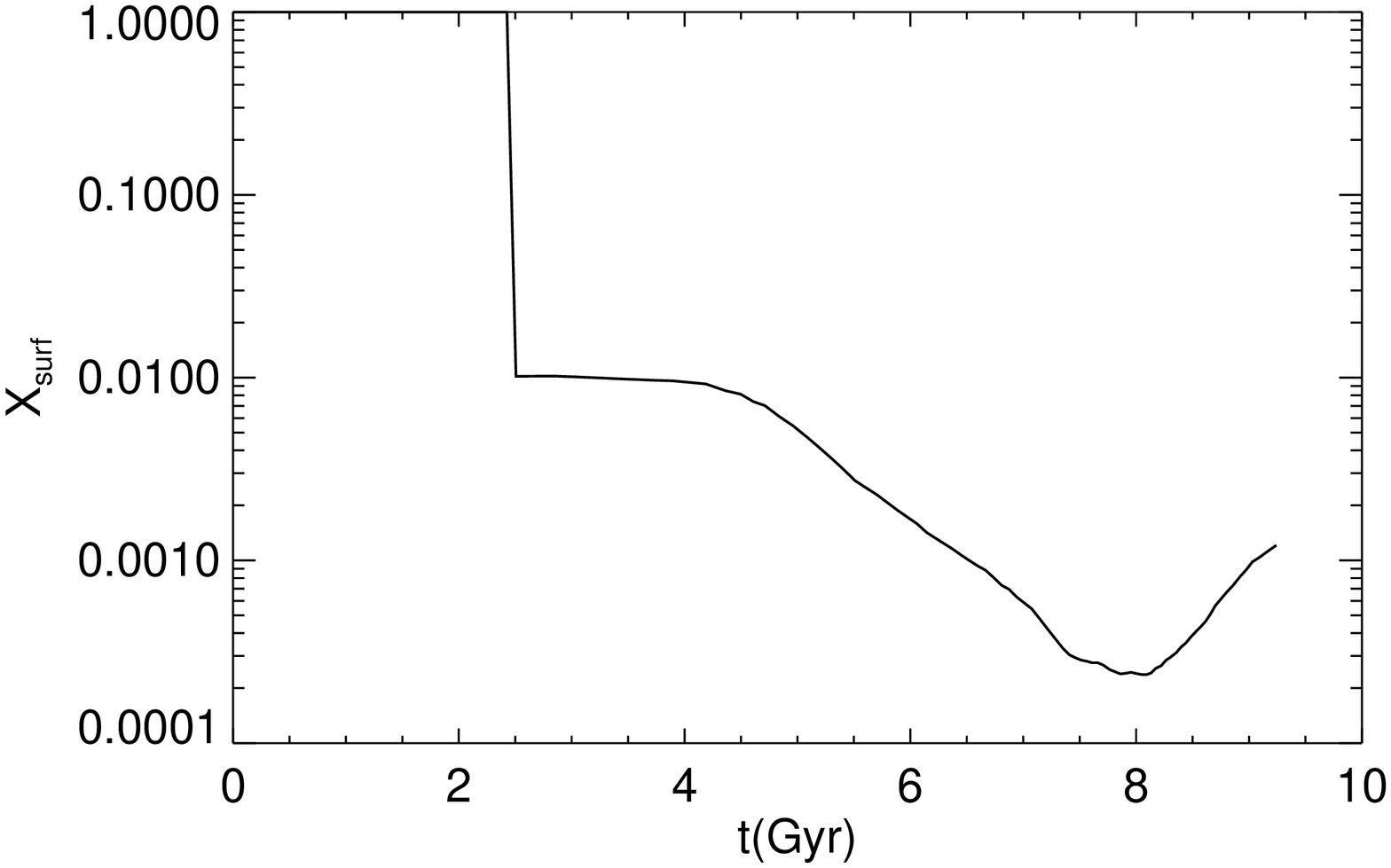}
\caption{An example of the cooling curves and chemical evolution curves of $0.6M_\sun$ WDs in \citet{2011MNRAS.413.2827C} (Figures~11 and~15, \copyright~2011 RAS).  UPPER: The cooling curve of the convective mixing WD with $m_H=10^{-8}M_\sun$ (solid line), that of the thick hydrogen layer WD (dashed line), and that of the pure helium WD (dash-dotted line).  LOWER: Chemical evolution curve of the convective mixing WD.  In the other two WDs, no chemical
evolution takes place.}
\label{f2}
\end{figure}

\begin{figure}
\includegraphics[width=0.75\textwidth]{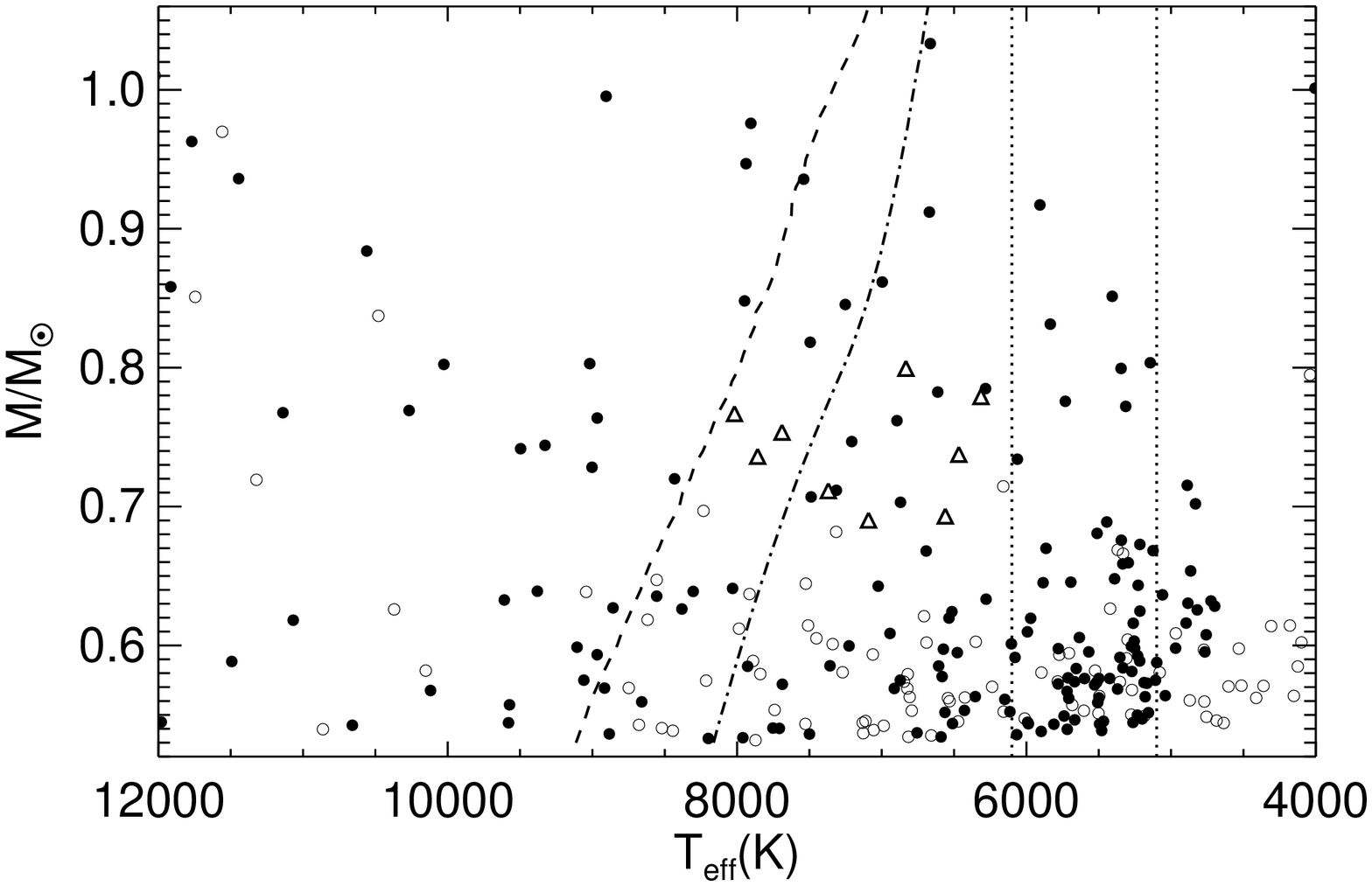}
\includegraphics[width=0.75\textwidth]{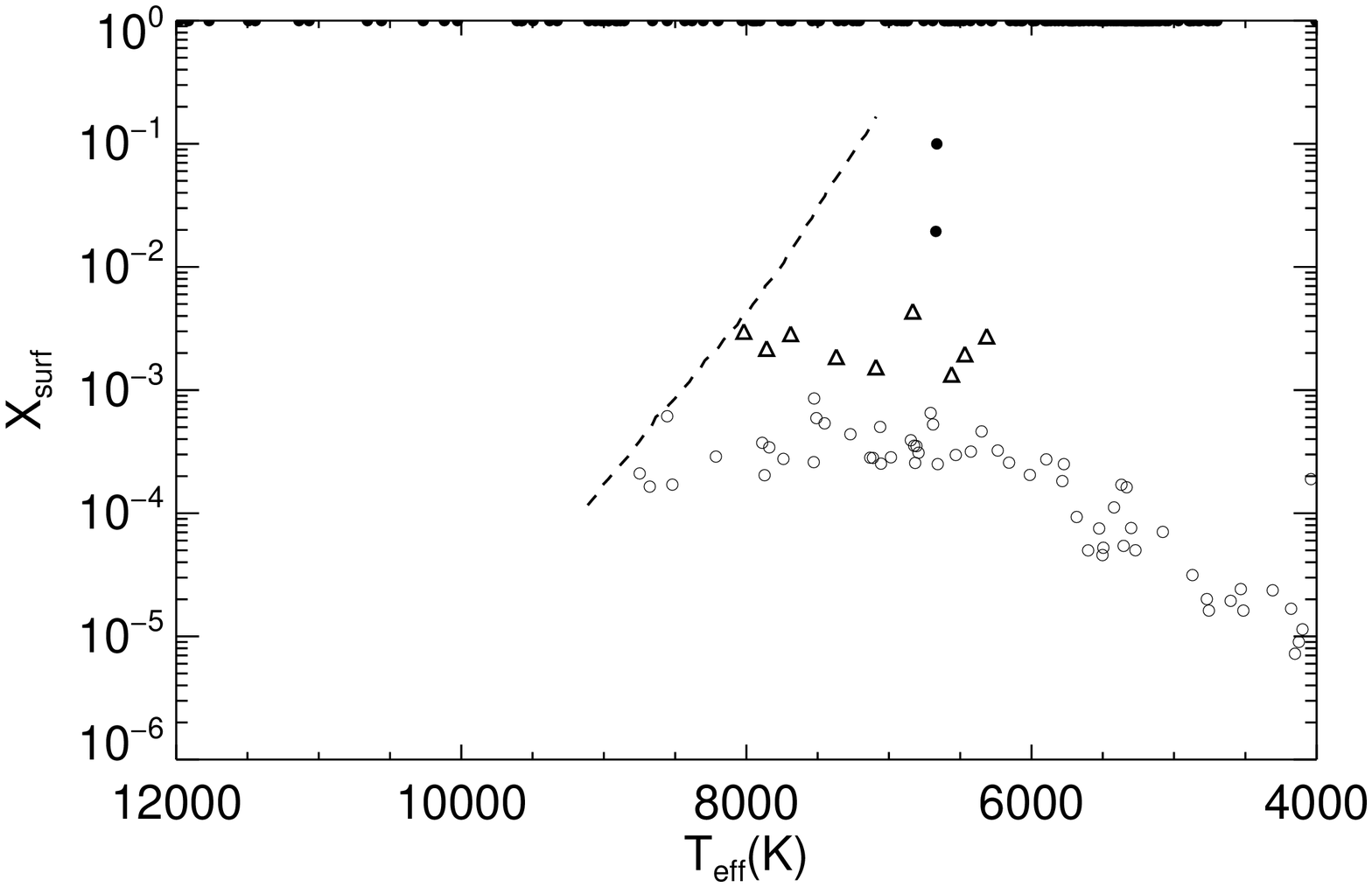}
\caption{Distribution of our representative Monte Carlo WD sample in two different parameter spaces.  The use of symbols is identical to that in Figure~\ref{f1}.  UPPER: Distribution in the $T_{eff}$--$M_{WD}$ space.  The dash-dotted line marks the $T_{eff}$ where convective mixing first occurs.  The dashed line marks the locus where the newly-mixed WDs converge.  The two dotted lines bracket a region of low non-DA ratio, corresponding to the ``non-DA gap''.  We note the $T_{eff}$ evolution of WD is non-monotonic when convective mixing occurs.  LOWER:  Distribution in $T_{eff}$--$X_{surf}$ space.  The dashed line marks the locus where the WDs converge after convective mixing.  The locus where convective mixing first occurs coincides with the upper edge of the plot and is therefore not explicitly present.  The thick hydrogen layer DA stars remain on the upper edge throughout their lifetime.  The pure helium WDs do not appear as they are located at infinity on this log scale plot.  The WDs cool faster after convective mixing
and are able to reach the low $T_{eff}$ end of the plot in large quantities within the given Galactic age.}
\label{f3}
\end{figure}

\begin{figure}
\includegraphics[width=0.8\textwidth]{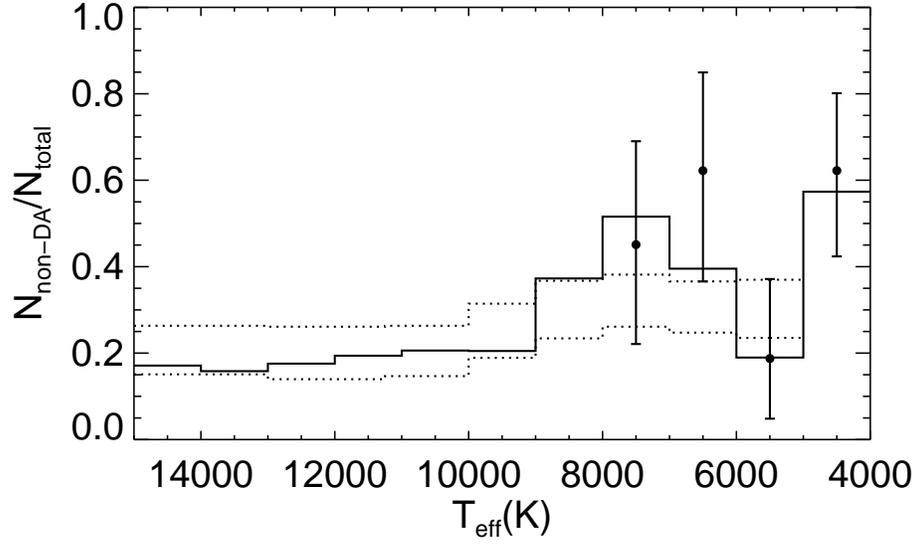}
\caption{Comparison of the non-DA to total WD ratio from our numerical work (solid line) to the observational data (dotted lines: \citet{2008ApJ...672.1144T}, filled circles: \citet{1998ApJ...497..294L}).  The binning intervals we used to compile the data in \citet{1998ApJ...497..294L} are $[4000K,5000K]$, $[5000K,6000K]$\dots, etc., to conform with the data of \citet{2008ApJ...672.1144T}.}
\label{f4}
\end{figure}

\begin{figure}
\includegraphics[width=0.8\textwidth]{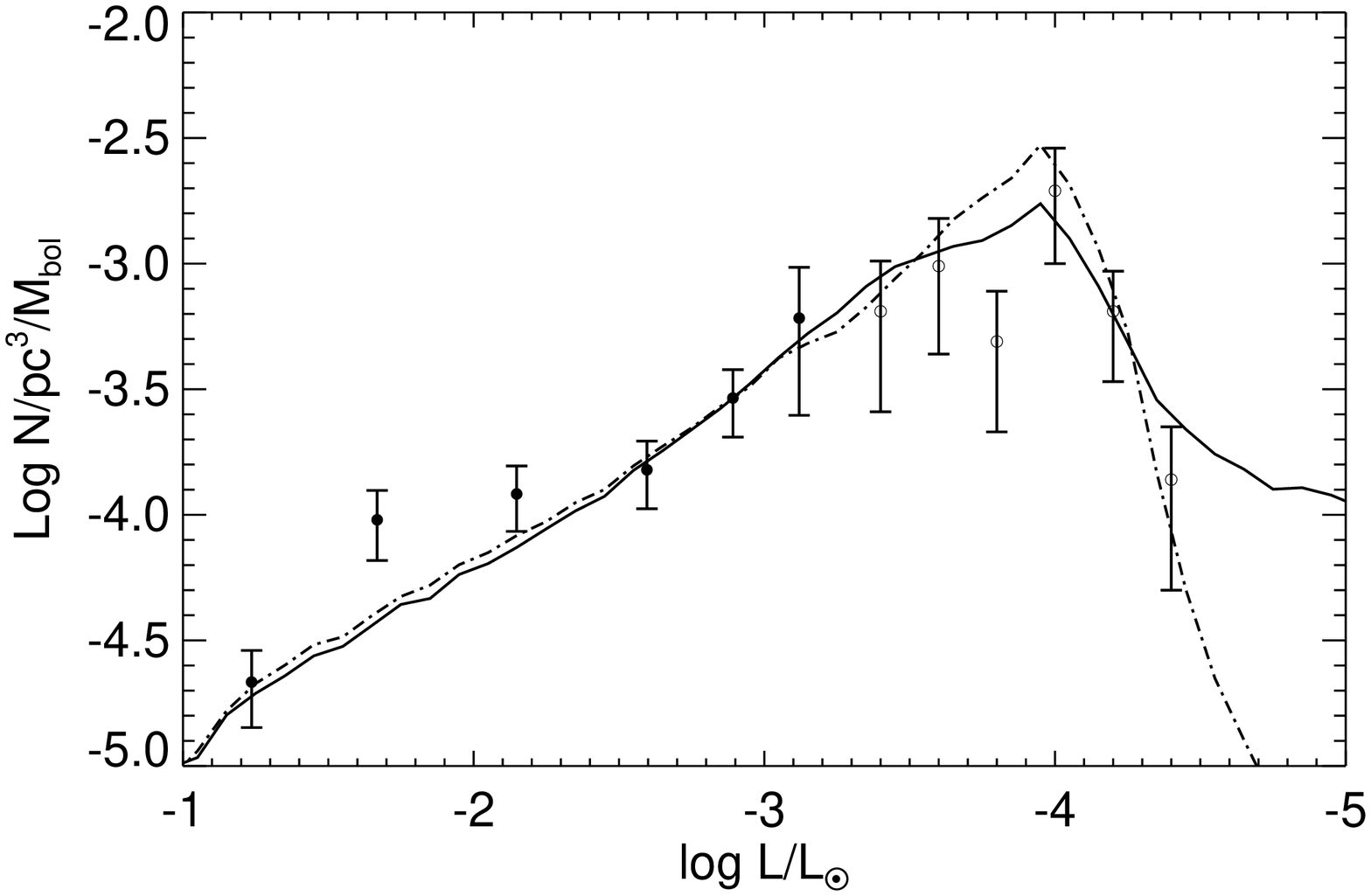}
\includegraphics[width=0.8\textwidth]{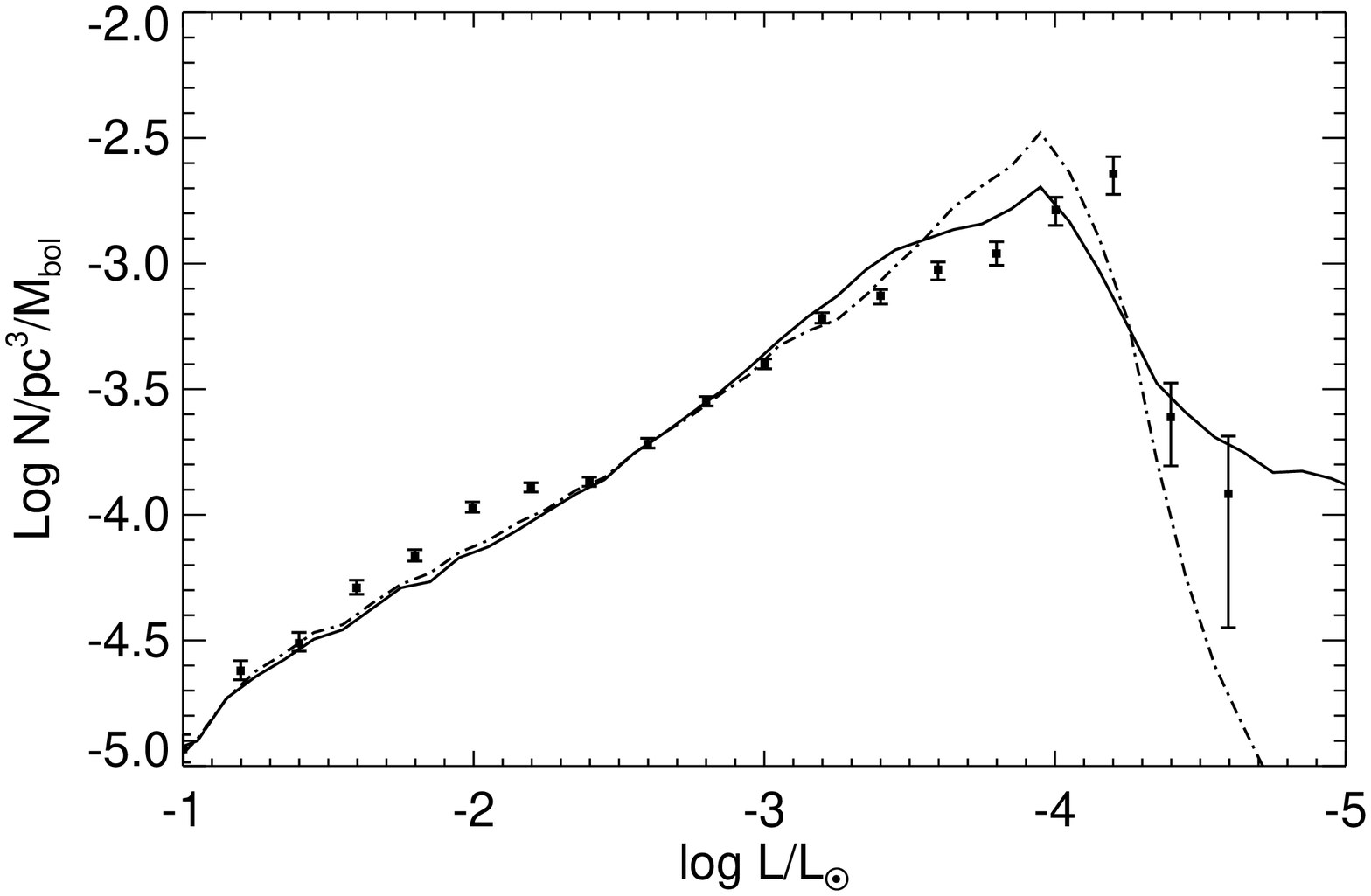}
\caption{UPPER: Theoretical CM-WDLF (solid line) and TH-WDLF (dash-dotted line), together with observational WDLF from \citet{1988ApJ...332..891L} (filled circle) and \citet{1998ApJ...497..294L} (hollow circle).  LOWER: Same theoretical construction compared with observational WDLF from SDSS \citep{2006AJ....131..571H}.}
\label{f5}
\end{figure}

Spectral evolution is important to WD Cosmochronology because the cooling rate of a WD is significantly affected by the atmospheric conditions (cf. Figure~\ref{f2}).  The nature of the faint tail of the observational WDLF based on the SDSS sample could be indicative of the problems faced by theoretical WDLF made with only thick hydrogen layer WDs.
    
Convective mixing, the non-DA gap, and WD Cosmochronology are closely related, in that advances in each topic will bring new insights to the others.  While this study has demonstrated the relation between them and established the consistency within the available observational results, works on each front, both theoretically and observationally, are not complete.
To that end, a spectroscopically confirmed WD sample larger than that of LDM/LRB is in need.  Although the sample from SDSS represents an important advancement, it suffers from the selection effects and halo WD contamination.  The results from future WD surveys will tell us more about the history of our Galaxy.

\acknowledgements
The authors acknowledge D. E. Winget, M. H. Montgomery, S. O. Kepler, and J. Adamczak for discussion and helpful comments. 
\bibliographystyle{apj}

\begin{thebibliography}{34}
\expandafter\ifx\csname natexlab\endcsname\relax\def\natexlab#1{#1}\fi

\bibitem[{{Althaus} \& {Benvenuto}(1997)}]{1997ApJ...477..313A}
{Althaus}, L.~G., \& {Benvenuto}, O.~G. 1997, \apj, 477, 313

\bibitem[{{Baglin} \& {Vauclair}(1973)}]{1973A&A....27..307B}
{Baglin}, A., \& {Vauclair}, G. 1973, \aap, 27, 307

\bibitem[{{Bedin} {et~al.}(2009){Bedin}, {Salaris}, {Piotto}, {Anderson},
  {King}, \& {Cassisi}}]{2009ApJ...697..965B}
{Bedin}, L.~R., {Salaris}, M., {Piotto}, G., {et~al.} 2009, \apj, 697, 965

\bibitem[{{Bedin} {et~al.}(2005){Bedin}, {Salaris}, {Piotto}, {King},
  {Anderson}, {Cassisi}, \& {Momany}}]{2005ApJ...624L..45B}
---. 2005, \apjl, 624, L45

\bibitem[{{Bergeron} {et~al.}(2001){Bergeron}, {Leggett}, \&
  {Ruiz}}]{2001ApJS..133..413B}
{Bergeron}, P., {Leggett}, S.~K., \& {Ruiz}, M.~T. 2001, \apjs, 133, 413

\bibitem[{{Bergeron} {et~al.}(1997){Bergeron}, {Ruiz}, \&
  {Leggett}}]{1997ApJS..108..339B}
{Bergeron}, P., {Ruiz}, M.~T., \& {Leggett}, S.~K. 1997, \apjs, 108, 339

\bibitem[{{Bergeron} {et~al.}(1990){Bergeron}, {Wesemael}, {Fontaine}, \&
  {Liebert}}]{1990ApJ...351L..21B}
{Bergeron}, P., {Wesemael}, F., {Fontaine}, G., \& {Liebert}, J. 1990, \apjl,
  351, L21

\bibitem[{{Chabrier} {et~al.}(2000){Chabrier}, {Brassard}, {Fontaine}, \&
  {Saumon}}]{2000ApJ...543..216C}
{Chabrier}, G., {Brassard}, P., {Fontaine}, G., \& {Saumon}, D. 2000, \apj,
  543, 216

\bibitem[{{Chen} \& {Hansen}(2011)}]{2011MNRAS.413.2827C}
{Chen}, E.~Y., \& {Hansen}, B.~M.~S. 2011, \mnras, 413, 2827

\bibitem[{{D'Antona} \& {Mazzitelli}(1990)}]{1990ARA&A..28..139D}
{D'Antona}, F., \& {Mazzitelli}, I. 1990, \araa, 28, 139

\bibitem[{{Fontaine} {et~al.}(2001){Fontaine}, {Brassard}, \&
  {Bergeron}}]{2001PASP..113..409F}
{Fontaine}, G., {Brassard}, P., \& {Bergeron}, P. 2001, \pasp, 113, 409

\bibitem[{{Garc{\'{\i}}a-Berro} {et~al.}(2008){Garc{\'{\i}}a-Berro}, {Althaus},
  {C{\'o}rsico}, \& {Isern}}]{2008ApJ...677..473G}
{Garc{\'{\i}}a-Berro}, E., {Althaus}, L.~G., {C{\'o}rsico}, A.~H., \& {Isern},
  J. 2008, \apj, 677, 473

\bibitem[{{Hansen}(1999)}]{1999ApJ...520..680H}
{Hansen}, B.~M.~S. 1999, \apj, 520, 680

\bibitem[{{Hansen} {et~al.}(2002){Hansen}, {Brewer}, {Fahlman}, {Gibson},
  {Ibata}, {Limongi}, {Rich}, {Richer}, {Shara}, \&
  {Stetson}}]{2002ApJ...574L.155H}
{Hansen}, B.~M.~S., {Brewer}, J., {Fahlman}, G.~G., {et~al.} 2002, \apjl, 574,
  L155

\bibitem[{{Hansen} {et~al.}(2007){Hansen}, {Anderson}, {Brewer}, {Dotter},
  {Fahlman}, {Hurley}, {Kalirai}, {King}, {Reitzel}, {Richer}, {Rich}, {Shara},
  \& {Stetson}}]{2007ApJ...671..380H}
{Hansen}, B.~M.~S., {Anderson}, J., {Brewer}, J., {et~al.} 2007, \apj, 671, 380

\bibitem[{{Harris} {et~al.}(2006){Harris}, {Munn}, {Kilic}, {Liebert},
  {Williams}, {von Hippel}, {Levine}, {Monet}, {Eisenstein}, {Kleinman},
  {Metcalfe}, {Nitta}, {Winget}, {Brinkmann}, {Fukugita}, {Knapp}, {Lupton},
  {Smith}, \& {Schneider}}]{2006AJ....131..571H}
{Harris}, H.~C., {Munn}, J.~A., {Kilic}, M., {et~al.} 2006, \aj, 131, 571

\bibitem[{{Hurley} {et~al.}(2000){Hurley}, {Pols}, \&
  {Tout}}]{2000MNRAS.315..543H}
{Hurley}, J.~R., {Pols}, O.~R., \& {Tout}, C.~A. 2000, \mnras, 315, 543

\bibitem[{{Koester}(1976)}]{1976A&A....52..415K}
{Koester}, D. 1976, \aap, 52, 415

\bibitem[{{Kroupa}(2001)}]{2001MNRAS.322..231K}
{Kroupa}, P. 2001, \mnras, 322, 231

\bibitem[{{Lamb} \& {van Horn}(1975)}]{1975ApJ...200..306L}
{Lamb}, D.~Q., \& {van Horn}, H.~M. 1975, \apj, 200, 306

\bibitem[{{Leggett} {et~al.}(1998){Leggett}, {Ruiz}, \&
  {Bergeron}}]{1998ApJ...497..294L}
{Leggett}, S.~K., {Ruiz}, M.~T., \& {Bergeron}, P. 1998, \apj, 497, 294

\bibitem[{{Liebert} {et~al.}(1988){Liebert}, {Dahn}, \&
  {Monet}}]{1988ApJ...332..891L}
{Liebert}, J., {Dahn}, C.~C., \& {Monet}, D.~G. 1988, \apj, 332, 891

\bibitem[{{Malo} {et~al.}(1999){Malo}, {Wesemael}, \&
  {Bergeron}}]{1999ApJ...517..901M}
{Malo}, A., {Wesemael}, F., \& {Bergeron}, P. 1999, \apj, 517, 901

\bibitem[{{Mestel}(1952)}]{1952MNRAS.112..583M}
{Mestel}, L. 1952, \mnras, 112, 583

\bibitem[{{Michaud} \& {Fontaine}(1984)}]{1984ApJ...283..787M}
{Michaud}, G., \& {Fontaine}, G. 1984, \apj, 283, 787

\bibitem[{{Muchmore}(1984)}]{1984ApJ...278..769M}
{Muchmore}, D. 1984, \apj, 278, 769

\bibitem[{{Renedo} {et~al.}(2010){Renedo}, {Althaus}, {Miller Bertolami},
  {Romero}, {C{\'o}rsico}, {Rohrmann}, \&
  {Garc{\'{\i}}a-Berro}}]{2010ApJ...717..183R}
{Renedo}, I., {Althaus}, L.~G., {Miller Bertolami}, M.~M., {et~al.} 2010, \apj,
  717, 183

\bibitem[{{Salaris} {et~al.}(2000){Salaris}, {Garc{\'{\i}}a-Berro}, {Hernanz},
  {Isern}, \& {Saumon}}]{2000ApJ...544.1036S}
{Salaris}, M., {Garc{\'{\i}}a-Berro}, E., {Hernanz}, M., {Isern}, J., \&
  {Saumon}, D. 2000, \apj, 544, 1036

\bibitem[{{Schmidt}(1975)}]{1975ApJ...202...22S}
{Schmidt}, M. 1975, \apj, 202, 22

\bibitem[{{Shipman}(1972)}]{1972ApJ...177..723S}
{Shipman}, H.~L. 1972, \apj, 177, 723

\bibitem[{{Tremblay} \& {Bergeron}(2008)}]{2008ApJ...672.1144T}
{Tremblay}, P.-E., \& {Bergeron}, P. 2008, \apj, 672, 1144

\bibitem[{{Winget} {et~al.}(1987){Winget}, {Hansen}, {Liebert}, {van Horn},
  {Fontaine}, {Nather}, {Kepler}, \& {Lamb}}]{1987ApJ...315L..77W}
{Winget}, D.~E., {Hansen}, C.~J., {Liebert}, J., {et~al.} 1987, \apjl, 315, L77

\bibitem[{{Wood}(1992)}]{1992ApJ...386..539W}
{Wood}, M.~A. 1992, \apj, 386, 539

\bibitem[{{Zhao} {et~al.}(2012){Zhao}, {Oswalt}, {Willson}, {Wang}, \&
  {Zhao}}]{2012ApJ...746..144Z}
{Zhao}, J.~K., {Oswalt}, T.~D., {Willson}, L.~A., {Wang}, Q., \& {Zhao}, G.
  2012, \apj, 746, 144

\end{thebibliography}

\newpage
\clearpage

\end{document}